\documentclass[aps,prl,reprint,groupedaddress]{revtex4-2}

\usepackage{graphicx}
\usepackage{bm}
\usepackage{amsmath,amsthm,amssymb,amscd}
\usepackage[colorlinks,linkcolor=blue, urlcolor=blue, anchorcolor=blue, citecolor=blue]{hyperref}
\DeclareMathAlphabet{\altmathcal}{OMS}{cmsy}{m}{n}
\usepackage{mathptmx}
\usepackage{microtype}
\usepackage{xcolor}
\newcommand{\cor}[1]{{\color{black} #1}}

\usepackage[normalem]{ulem}

\begin{document}


\title{Metastability-Induced Solid-State Quantum Batteries for Powering Microwave Quantum Electronics}


\author{Yuanjin Wang}
\affiliation{Center for Quantum Technology Research and Key Laboratory of Advanced Optoelectronic Quantum Architecture and Measurements (MOE), School of Physics, Beijing Institute of Technology, Beijing 100081, China}

\author{Hao Wu}
\email[]{hao.wu@bit.edu.cn}
\affiliation{Center for Quantum Technology Research and Key Laboratory of Advanced Optoelectronic Quantum Architecture and Measurements (MOE), School of Physics, Beijing Institute of Technology, Beijing 100081, China}

\author{Qing Zhao}
\email[]{qzhaoyuping@bit.edu.cn}
\affiliation{Center for Quantum Technology Research and Key Laboratory of Advanced Optoelectronic Quantum Architecture and Measurements (MOE), School of Physics, Beijing Institute of Technology, Beijing 100081, China}


\date{\today}

\begin{abstract}
Metastability is ubiquitous in diverse complex systems. In open quantum systems, metastability offers protection against dissipation and decoherence, yet its application in quantum batteries remains unexplored. We propose a solid-state open quantum battery where metastable states enable stable superextensive charging without complicated protocols and energy storage with extended lifetime. Using a realistic organic maser platform, we show the controllable manner of the work extraction from the quantum battery, which can be exploited for on-demand coherent microwave emission at room temperature. These results not only demonstrate the usefulness of metastability for developing the quantum batteries robust against energy losses, but also provide a paradigm of the practical quantum device powered up by quantum batteries.

\end{abstract}


\maketitle









In nature, due to intrinsic complexity, systems not only settle into stable configurations but also frequently manifest in non-equilibrium states. Metastability, a state in which a system remains in a non-equilibrium condition for an extended period before relaxing to true stationarity\cite{penrose1971rigorous}, is ubiquitous in a variety of complex classical systems\cite{bovier2016metastability}, such as spin glasses\cite{binder1986spin,fisher1988equilibrium,newman2003ordering}, large (bio-)molecules\cite{grassberger1998monte, gliko2007metastable,chodera2007automatic} and genetic populations\cite{van2000metastable,dawson2014spatial,kelso2012multistability}. The exploitation of metastability is of considerable practical interest that has driven recent advances in the realms of next-generation technological material designs\cite{sun2016thermodynamic}, protein crystallography\cite{khurshid2014porous} and cryptography\cite{sivaraman2020metastability}.

The studies on metastability within the aforementioned classical systems predominantly focus on its feature of \textit{randomness}\cite{bovier2016metastability}, while in the quantum regime, because the real-world quantum systems are generically complex (e.g., many-body) and unavoidably coupled to the environment (i.e., open), leading to dissipation and decoherence\cite{breuer2002theory,gardiner2004quantum}, the research emphasis is shifting towards the \textit{protective} characteristic of quantum metastability arising from the disjoint states, noiseless subsystems and decoherence-free subspaces involved in the mainfold of metastable states\cite{macieszczak2016towards}. Although recent years have witnessed burgeoning interest in understanding quantum metastability found in a wide range of open quantum many-body systems\cite{letscher2017bistability,souto2017quench,macieszczak2017metastable,hruby2018metastability,blass2018quantum,jin2024theory}, the explicit applications of quantum metastability are currently limited within the field of quantum information processing\cite{ma2023high,debry2023experimental,shi2024long}.

The quantum battery is a quantum system capable of both energy storage and extraction, representing a straightforward application scenario within the field of quantum thermodynamics\cite{Goold_2016, Quach2023,campaioli2024colloquium}. The utilization of quantum effects, such as coherence\cite{PhysRevLett.125.040601,PhysRevLett.122.047702,PhysRevB.102.245407,PhysRevA.107.012207,PhysRevLett.129.130602,PhysRevE.102.052109,Latune2019} and entanglement\cite{PhysRevE.87.042123,PhysRevLett.118.150601,PhysRevLett.120.117702,PhysRevLett.111.240401,PhysRevLett.131.060402,PhysRevA.107.022215,PhysRevB.105.115405,PhysRevLett.128.140501,PhysRevB.104.245418,PhysRevA.106.032212,PhysRevResearch.2.023113,PhysRevA.104.L030402,PhysRevA.110.022425,PhysRevLett.127.100601,PhysRevB.100.115142}, facilitates the achievement of performance surpassing that of classical batteries, making it a subject of extensive interest since its inception\cite{PhysRevE.87.042123}. To advance the practical realization, there has been a growing body of research focused on the theory 
and experiments 
of \textit{open quantum batteries}\cite{campaioli2024colloquium,PhysRevLett.132.210402}, in which, addressing inevitable energy loss caused by its dissipative nature has become a crucial task for device implementation. The use of dark/subradiant\cite{freedhoff1967theory,stroud1972superradiant} states has been proposed\cite{PhysRevApplied.14.024092} to tackle this issue of quantum batteries with only few spins and at extremely low temperature, leaving its scalability and feasibility at high temperature ambiguous. Alternatively, Campaioli \textit{et al.}\cite{campaioli2024colloquium} suggested to exploit the protective nature of metastability for mitigating energy loss in open quantum batteries. Indeed, the molecular photoexcited triplet states, which are normally metastable and longer-lived compared to the corresponding singlet states\cite{mcclure1949triplet}, have been recently proposed\cite{tibben2024extending}.
However, the (dis)charging performance of the proposed quantum battery remains elusive and the optimal energy-transfer mechanism awaits suitable platforms to be realized.

In this letter, we propose a new implementation of the solid-state open quantum battery in which the metastable states possessing fine structures act as the main body (i.e. the battery) for energy storage and extraction instead of serving as the intermediates for extending the lifetime of a 'master' battery comprised by the stable (singlet) states\cite{tibben2024extending}. In addition to the prolonged energy-storage lifetime, the metastability-induced quantum battery shows the intrinsic advantage for achieving stable charging without the disconnection of the charger or complex stabilization protocols. We show such a quantum battery is experimentally feasible with the realistic room-temperature organic maser platform\cite{oxborrow2012room}. Intriguingly, the quantum battery displays a superextensive charging-power scaling that beats the limit of the common collective charging protocol\cite{PhysRevLett.128.140501}. Moreover, we investigate the influence of \cor{external} driving fields on charging and work-extraction behaviors, which provides a theoretical foundation for applying the quantum battery in powering the microwave quantum electronics, e.g., quantum oscillators\cite{wang2024tailoring}.

\begin{figure}
	\includegraphics[]{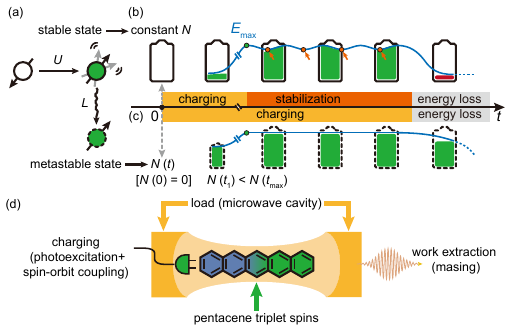}
    \caption{\label{fig1} (a) A quantum battery composed by $N$ stable quantum systems is collectively charged via unitary operations $U$ and subsequently evolves via dissipative processes $L$ into a metastable phase where the number of newly emerged quantum systems $N(t)$ is time dependent. (b) A conventional quantum battery requires precise charging and stabilization protocols and self-discharges quickly. (c) A metastability-induced quantum battery has a time-varying energy capacity, stable charging behaviour and long energy-storage lifetime. \cor{(d) Practical implementation and usage of the proposed quantum battery based on a pentacene maser system.}}
\end{figure}

The proposed quantum battery is constructed by the non-degenerate metastable spin states obtained through a dissipative process, e.g., the spin-orbit coupling, following the primary collective charging of the stable quantum systems into the excited states, as shown in Fig.\hyperref[fig1]{1(a)}. The dissipative process shows the anisotropic property that leads to the population inversion in the metastable states, thus,  charging the quantum battery. Such a process is typically found in organic doped systems\cite{van2017candidate} due to the weak spin-orbit coupling\cite{de1967phosphorescence} and gives rise to the singlet-triplet transitions, known as the intersystem crossing. In contrast to the conventional quantum batteries that maintain a fixed number during charging [Fig.\hyperref[fig1]{1(b)}] , the metastability-induced quantum battery is generated and charged simultaneously, thus its size and the capacity would increase during the charging time before reaching the maximum [Fig.\hyperref[fig1]{1(c)}]. Due to the dissipative process that decouples the quantum battery from the primary charging field, the system is immune to the oscillatory behavior during charging\cite{santos2019stable}, while the conventional quantum batteries normally require unplugging the charger or precise charging and stabilization protocols to avoid the energy fluctuation\cite{campaioli2024colloquium}. Furthermore, owing to the protective nature of metastability, the proposed quantum battery offers the advantage of the  extended energy storage duration to mitigate the issue of rapid self-discharging encountered by its conventional counterparts.

\begin{figure*}
	\includegraphics{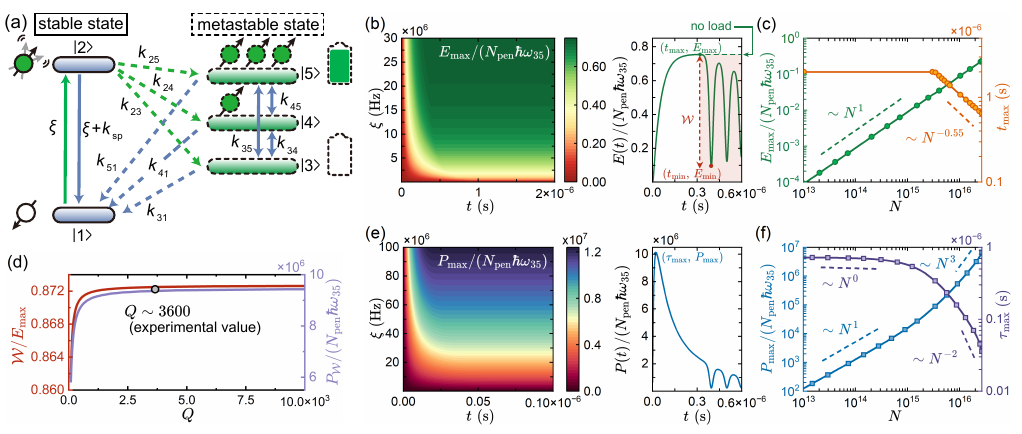}
	\caption{\label{fig2} Mechanism and performance of the metastability-induced solid-state quantum battery. (a) General mechanism illustrating the quantum battery based on the \cor{pentacene's} metastable triplet states formed via the intersystem crossing from the photoexcited singlet states. 
    \cor{See Supplemental Material \cite{Supplemental} for the complete parameter setting}. (b), (e) Dependence of the quantum battery's maximum stored energy $E_{\mathrm{max}}$ and maximum charging power $P_{\mathrm{max}}$ on the charging time $t$ and photoexcitation rate $\xi$. The definitions of $E_{\mathrm{max}}$ and $P_{\mathrm{max}}$ together with their reaching times $t_{\mathrm{max}}$ and $\tau_{\mathrm{max}}$ are shown with the time evolution of the stored energy $E(t)$ (solid green line) and charging power $P(t)$ (solid blue line), where $\xi=6.2\times10^7$ Hz (corresponding to a photoexcitation power of $2\times10^4$ W). \cor{The dashed double arrow defines the maximum extractable work $\altmathcal{W}$ of the quantum battery in realistic operation where the load, i.e., a microwave cavity presents for the work extraction (red shadow). Without the load, the energy is stored for a long duration (dashed green line).} (d) \cor{Dependence of the ratio of the maximum extractable work to the maximum energy ($\altmathcal{W} / E_{\mathrm{max}}$) and the maximum discharging power $P_\altmathcal{W}$ on the cavity quality factor $Q$. The gray circle indicates the condition where the quantum battery can achieve with the existing experimental setup).} (c), (f) Scaling of $E_{\mathrm{max}}$ and $P_{\mathrm{max}}$ as well as the reaching times $t_{\mathrm{max}}$ and $\tau_{\mathrm{max}}$ with the size of the quantum battery $N$. 
 }
\end{figure*}

To demonstrate the feasibility of achieving the proposed quantum battery and its practical usage for powering loads/devices, we employ a realistic platform, so-called the pentacene maser system\cite{oxborrow2012room}, which enables the straightforward implementation \cor{in Earth's field} in solid states at room temperature. As illustrated in \cor{Fig.\hyperref[fig1]{1(d)}}, the system is constituted by the pentacene molecules (i.e. the battery) doped in a \textit{p}-terphenyl crystal housed in a microwave cavity (i.e. the load). There are two stages for charging the pentacene quantum battery \cor{shown in Fig.\hyperref[fig2]{2(a)}}. The primary charging field is the optical pumping of the pentacene molecules in its ground singlet state $|1\rangle$ to the excited singlet state $|2\rangle$. Subsequently, the intersystem crossing acts as the dissipative process that introduces the long-lived metastable triplet configuration of the quantum battery, in which the triplet sublevels ($|5\rangle$ to $|3\rangle$), with lifetimes ($\sim\mu$s to ms\cite{wu2019unraveling}) much longer than the singlet states ($\sim$ns\cite{takeda2002zero}), exhibit the population inversion with an instantaneous ratio of $0.76:0.16:0.08$\cite{sloop1981electron}, thus charging the battery via selectively populating the state $|5\rangle$. The energy stored in the quantum battery can be extracted to the loaded microwave cavity via the stimulated emission of coherent microwave photons (i.e. masing from $|5\rangle$ to $|3\rangle$).

To depict the dynamics of the pentacene quantum battery, we present the quantum master equation at time $t$ for the density operator $\hat{\rho}$\cite{Wu2024,PhysRevA.99.022302}:
\begin{equation}\label{eq:eq1}
\begin{aligned}
 \partial_\mathrm{t} \hat{\rho}=&-\frac{i}{\hbar}\left[\hat{H}_{\mathrm{bat}}+\hat{H}_\mathrm{m}+\hat{H}_{\mathrm{m-bat}}+\hat{H}_{\mathrm{drive}}, \hat{\rho}\right] \\
& +\frac{1}{2}\xi \sum_\mathrm{k} \mathcal{D}\left[\hat{\sigma}_\mathrm{k}^{\mathrm{21}}\right] \hat{\rho}+\frac{1}{2}\left(\xi+k_{\mathrm{sp}}\right) \sum_\mathrm{k} \mathcal{D}\left[\hat{\sigma}_\mathrm{k}^{\mathrm{12}}\right] \hat{\rho} \\
& +\sum_\mathrm{k}\left(\sum_{\mathrm{i=3,4,5}}\frac{1}{2} k_{\mathrm{2i}} \mathcal{D}\left[\hat{\sigma}_\mathrm{k}^{\mathrm{i2}}\right] \hat{\rho}+\sum_{\mathrm{i=3,4,5}}\frac{1}{2} k_{\mathrm{i1}} \mathcal{D}\left[\hat{\sigma}_\mathrm{k}^{\mathrm{1i}}\right] \hat{\rho}\right) \\
& +\sum_\mathrm{k} \sum_{\mathrm{i, j=3,4,5 ; i \neq j}}\left(\frac{1}{2}k_{\mathrm{ij}} \mathcal{D}\left[\hat{\sigma}_\mathrm{k}^{\mathrm{ji}}\right] \hat{\rho}\right) \\
& +\sum_\mathrm{k} \sum_{\mathrm{i, j=3,4,5 ; i < j}}\left(\frac{1}{4} \chi_{\mathrm{ij}} \mathcal{D}\left[\hat{\sigma}_\mathrm{k}^{\mathrm{jj}}-\hat{\sigma}_\mathrm{k}^{\mathrm{ii}}\right] \hat{\rho}\right) \\
& +\frac{\kappa}{2}\left[\left( n^{\mathrm{th}}_{\mathrm{m}} +1 \right)\mathcal{D}[\hat{a}] \hat{\rho}+n^{\mathrm{th}}_{\mathrm{m}} \mathcal{D}[\hat{a}^{\dagger}] \hat{\rho}\right],
\end{aligned}
\end{equation}
where $\mathcal{D}[\hat{\altmathcal{O}}]\hat{\rho}=2\hat{\altmathcal{O}}\hat{\rho}\hat{\altmathcal{O}}^\dagger-\hat{\altmathcal{O}}^\dagger\hat{\altmathcal{O}}\hat{\rho}-\hat{\rho}\hat{\altmathcal{O}}^\dagger\hat{\altmathcal{O}}$ is the Lindblad superoperator. $\hat{H}_{\mathrm{bat}}=\frac{1}{2}\hbar \cor{\Delta_{\mathrm{B}}} \sum_{\mathrm{k=1}}^{N_{\mathrm{pen}}} \hat{\sigma}_\mathrm{k}^{\mathrm{z}} $ \cor{describes the quantum batteries with $\Delta_{\mathrm{B}}=\omega_{\mathrm{35}}-\omega_{\mathrm{d}}$ (transition frequency $\omega_{\mathrm{35}}$ and external diving field frequency $\omega_{\mathrm{d}}$}) \cite{PhysRevLett.128.253601}.
Here, $\hat{\sigma}_\mathrm{k}^{\mathrm{z}}=\left(\hat{\sigma}_\mathrm{k}^{\mathrm{55}}-\hat{\sigma}_\mathrm{k}^{\mathrm{33}} \right) $, $\hbar$ is the reduced Planck constant, $\textrm{k}$ and $N_{\mathrm{pen}}$ indicate the individual and total number of pentacene molecules. The operators $\hat{\sigma}_\mathrm{k}^{\mathrm{ij}}=\left|i_\mathrm{k}\right\rangle\left\langle j_\mathrm{k}\right|$ represent transition operators for $i \neq j$ and projection operators for $i = j$. The photon creation and annihilation operators are denoted by $\hat{a}^{\dagger}$ and $\hat{a}$, respectively. $\hat{H}_\mathrm{m}=\hbar \cor{\Delta_{\mathrm{m}}} \hat{a}^{\dagger} \hat{a}$ characterizes the microwave cavity mode functioning as a load with \cor{$\Delta_{\mathrm{m}}=\omega_\mathrm{m}-\omega_{\mathrm{d}}$ and} frequency $\omega_\mathrm{m}$. $\hat{H}_{\mathrm{m-bat}}=\hbar \sum_\mathrm{k} g_{\mathrm{35}}\left(\hat{\sigma}_\mathrm{k}^{\mathrm{53}} \hat{a}+\hat{a}^{\dagger} \hat{\sigma}_\mathrm{k}^{\mathrm{35}}\right)$ describes the interaction between the pentacene quantum batteries and the microwave cavity with the coupling strength $g_{\mathrm{35}}$. $\hat{H}_{\mathrm{drive}}=\cor{\hbar} \sum_\mathrm{k} \sum_{\mathrm{i, j=3,4,5 ; i < j}} \Omega_{\mathrm{ij}}\left(\hat{\sigma}_\mathrm{k}^{\mathrm{ij}}+\hat{\sigma}_\mathrm{k}^{\mathrm{ji}}\right)$ represents an \cor{external} driving field with the strength $\Omega_{\mathrm{ij}}$. \cor{The frequency of the external driving field $\omega_{\mathrm{d}}$ is always set to be resonant with the associated spin transitions ($\omega_{\mathrm{35}}$ or $\omega_{\mathrm{45}}$ herein) within the quantum battery.}

The second to sixth rows of Eq.\hyperref[eq:eq1]{(1)} respectively represent the optical pumping and the spontaneous emission with the rates $\xi, k_{\mathrm{sp}}$, the intersystem crossing with the rates $k_{\mathrm{2i}}, k_{\mathrm{i1}}(i=3,4,5)$ to and from the triplet states, the spin-lattice relaxation with rates $k_{\mathrm{34}} \approx k_{\mathrm{43}}, k_{\mathrm{35}} \approx k_{\mathrm{53}}, k_{\mathrm{45}} \approx k_{\mathrm{54}}$, the dephasing process with the rates $\chi_{\mathrm{35}}, \chi_{\mathrm{45}}, \chi_{\mathrm{34}}$, and the thermal emission and excitation of the microwave cavity (photon damping rate $\kappa$) with the thermal equilibrium photon number $n^{\mathrm{th}}_{\mathrm{m}}$. Based on the mean-field approach\cor{\cite{debnath2018lasing,zhang2021ultranarrow,Plankensteiner2022quantumcumulantsjl}}, $\partial_\mathrm{t}\langle\hat{\altmathcal{O}}\rangle=\operatorname{tr}\{\hat{\altmathcal{O}} \partial_\mathrm{t}\hat{\rho}\}$ is derived for obtaining the operator expectation values $\langle\hat{\altmathcal{O}}\rangle$ (see Supplemental Material \cite{Supplemental}).

We utilize the summation of the pentacene molecules in states $|5\rangle$ and $|3\rangle$ as the total number $N$ of the quantum batteries, \cor{that is $N=N_{\mathrm{pen}}\left(  \langle \hat{\sigma}_\mathrm{1}^{\mathrm{55}}(t)\rangle+  \langle \hat{\sigma}_\mathrm{1}^{\mathrm{33}}(t)\rangle \right)$,  with $\langle \hat{\sigma}_\mathrm{1}^{\mathrm{jj}}(t)\rangle$ the population in state $|j\rangle$ at time $t$.} 
The relatively small and stable population in $|4\rangle$ does not impact the analysis (see Supplemental Material \cite{Supplemental}). 
The total energy accumulated in the quantum battery can be expressed as 
$E(t)=N_{\mathrm{pen}}\hbar \omega_{\mathrm{35}} \langle \hat{\sigma}_\mathrm{1}^{\mathrm{55}}(t)\rangle$.
The charging power $P$ is directly derived from 
$P=N_{\mathrm{pen}}\hbar\omega_{35} \langle \hat{\sigma}_\mathrm{1}^{\mathrm{55}}(t)\rangle/t$.
We denote $t_{\mathrm{max}}$ as the time that the energy stored reaches its first maximum $E_{\mathrm{max}}$. 
And time $\tau_{\mathrm{max}} \in (0,t_{\mathrm{max}})$ corresponds the occurrence of maximum charging power $P_{\mathrm{max}}$. 

%

Figs.\hyperref[fig2]{2(b)} and \hyperref[fig2]{(e)} depict the charging performance of the pentacene quantum battery under different intensities $\xi$ of the pump laser. As the size of the quantum battery depends on the optical pumping, by varying $\xi$, we can analyze the scaling of the energy storage and charging dynamics with $N$. In Fig.\hyperref[fig2]{2(b)}, the small value of $\xi$ (below $5 \times 10^6$ Hz) fails to compensate the spontaneous emission of state $|2\rangle$ and the decay of the triplet states to state $|1\rangle$, thereby impeding efficient charging of the quantum battery. When $\xi$ reaches a sufficient strength (about $10^7$ Hz) to compensate these two processes, it enables stable charging of the quantum battery to $E_\textrm{max}$. \cor{And the energy-storage lifetime $\tau_\mathrm{s}$ is determined by the decay rate of the population in $|5\rangle$\cite{pirmoradian2019aging} without the load.
 $\tau_\mathrm{s}$ can reach 25 and 30 $\mu$s, respectively, depending on whether the charging stops (see the Supplemental Material \cite{Supplemental}). It is found that even the charging continues when $E_\textrm{max}$ is reached, the energy will not fluctuate, proving the stability of the quantum battery. The obtained storage lifetime is three orders of magnitude longer than that in state-of-the-art quantum batteries exploiting metastable triplet states for the extension of energy storage\cite{hymas2025experimental}.}

\cor{When the battery is loaded}, the maximum stored energy \cor{$E_\textrm{max}/(N_\textrm{pen}\hbar\omega_\textrm{35})\sim0.75$} can be extracted for work \cor{in form of masing} at a certain time, e.g., $\sim$300 ns, resulting in the collapse of the stored energy and charging power \cor{shown in Figs.\hyperref[fig2]{2(b)} and \hyperref[fig2]{(e)}. And the maximum extractable energy for work, $\altmathcal{W}/(N_\textrm{pen}\hbar\omega_\textrm{35})\sim0.65$, can be obtained via $\altmathcal{W}=E_\textrm{max}-E_\textrm{min}$, since the discharging to the minimum energy $E_\textrm{min}$ position is dominated by the masing process (see the Supplemental Material \cite{Supplemental}) where the extracted energy is available as electromagnetic work to power the load, that is, to oscillate the microwave cavity. The maximum discharging power is thus $P_\altmathcal{W}=(E_{\mathrm{max}}-E_{\mathrm{min}})/(t_{\mathrm{min}}-t_{\mathrm{max}})$. The efficiency of the work extraction can be evaluated by $\altmathcal{W}/E_\textrm{max}$\cite{PhysRevLett.127.100601}. As indicated in Fig.\hyperref[fig2]{2(d)}, using the parameters extracted from the experiments\cite{wu2020room}, $\altmathcal{W} / E_{\mathrm{max}}$ of the proposed quantum battery exceeds 0.87, which is greater than that obtained experimentally in a nitrogen vacancy center-based single-spin system\cite{PhysRevLett.133.180401}. Note that, due to the presence of dissipation during the work extraction, the parameter $\altmathcal{W}$ is not exactly equivalent to the concept of ergotropy\cite{A.E.Allahverdyan_2004}. We found that the larger quality factor ($Q$) of the load could enhance the efficiency as well as the discharging power, and remarkably, the available experimental setup has almost reached the optimum. Moreover}, arising from the collective coupling of the pentacene triplet spins with the microwave cavity mode\cite{Breeze2017}, the battery energy can be restored automatically with an underdamping feature as indicated by the time evolution results $E(t)$ and $P(t)$ in Figs.\hyperref[fig2]{2(b)} and \hyperref[fig2]{(e)}. The similar behaviors have been experimentally observed by the microwave-photon statistics of the pentacene masers\cite{Breeze2017,wu2020room}.



Fig.\hyperref[fig2]{2(c)} and \hyperref[fig2]{2(f)} illustrate the dependence of $t_{\mathrm{max}}, E_{\mathrm{max}}, \tau_{\mathrm{max}}$ and $P_{\mathrm{max}}$ on $N$. $E_\textrm{max}$ is proportional to $N$, while for $t_\textrm{max}$, once $N$ is increased beyond $2 \times 10^{15}$, $t_\textrm{max}$ starts to drop and reveals a scaling of $t_{\mathrm{max}} \propto N^{\mathrm{-0.55}}$, which is superior to the conventional collective charging protocol where $t_{\mathrm{max}} \propto N^{\mathrm{-0.5}}$\cite{PhysRevLett.120.117702}.
Meanwhile, fitting the variation of $P_\textrm{max}$ with $N$ gives the linear scaling ($\propto N$) like classical batteries when $N$ is below $10^{15}$. The larger size of the quantum battery is accompanied by a significant reduction in the charging time $\tau_{\mathrm{max}}$ ($\propto N^{\mathrm{-2}}$), leading to the scaling $P_{\mathrm{max}} \propto N^{\mathrm{3}}$, which is even higher than the most quadratic scaling in charging power ($\propto N^{2}$) predicted for the global charging operations\cite{PhysRevLett.128.140501}. We observe a similarity between the scalings of $\xi$ and $P_{\mathrm{max}}$ with $N$ (see Supplemental Material \cite{Supplemental}), indicating the pump intensity can be adjusted to surpass the conventional scaling limit of charging power.





\begin{figure}[t]
	\includegraphics[width=0.48\textwidth]{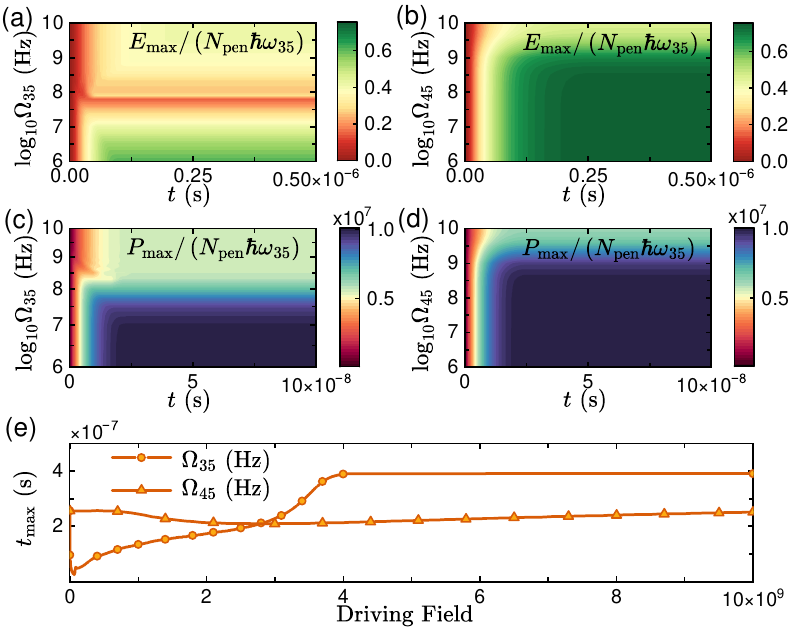}
\caption{\label{fig3} Effects of the driving fields on the quantum battery performance. (a)$-$(d) Dependence of $E_{\mathrm{max}}$ and $P_{\mathrm{max}}$ on the classical driving fields $\Omega_{\mathrm{35}}$ and $\Omega_{\mathrm{45}}$ as well as the charging time $t$. (e) Dependence of the time to achieve $E_{\mathrm{max}}$, i.e. $t_{\mathrm{max}}$ on $\Omega_{\mathrm{35}}$ and $\Omega_{\mathrm{45}}$. The results were computed under $\xi=6.2\times10^7$ Hz.}
\end{figure}

We also investigate the impact of the \cor{external} driving fields on the charging performance. As shown in Figs.\hyperref[fig3]{3(a)-(d)}, the increasing driving fields ($\Omega_{\mathrm{35}}$ and $\Omega_{\mathrm{45}}$) can lead to the reduction of $E_\textrm{max}$ and $P_\textrm{max}$. This can be explained by the fact that the initially prepared spin state of the quantum battery is inverted due to the selective populating process (i.e., the intersystem crossing), so the driving fields can diminish the population difference and favor the transitions from $|5\rangle$ to $|3\rangle$ and $|4\rangle$ introducing the discharging channels. And since the population difference between $|5\rangle$ and $|3\rangle$ is larger than that between \cor{$|5\rangle$ and $|4\rangle$}, the driving-field-induced discharging is more evident with $\Omega_{\mathrm{35}}$.  On the other hand, we note that \cor{compared to $E_\textrm{max}$,} $P_\textrm{max}$ is more robust against with the driving field $\Omega_{\mathrm{35}}$. This could be attributed to the effect of the driving fields on $t_\textrm{max}$ as shown in Fig.\hyperref[fig3]{3(e)}. As the charging rate is fixed and the driving fields are decoupled from the charging process, the reduced $E_\textrm{max}$ will require less time to reach and the \cor{minimum} $t_\textrm{max}$ \cor{obtained with} $\Omega_{\mathrm{35}}$ is about 1/3 of that obtained with $\Omega_{\mathrm{45}}$. \cor{The increase in $\Omega_{\mathrm{35}}$ prior to approximately $10^8$ Hz results in a sharp decline in $t_\textrm{max}$ matching the rapid decrease in $E_\textrm{max}$ as shown in Fig.\hyperref[fig3]{3(a)}. When $\Omega_{\mathrm{35}}$ exceeds $10^8$ Hz, the population ratio of $|5\rangle$ to $|3\rangle$ at $E_\textrm{max}$ approaches $1:1$ (see Supplemental Material \cite{Supplemental}), the decrease of $E_\textrm{max}$ becomes minor and the driving field facilitates rapid population transfer from $|5\rangle$ to $|3\rangle$, thereby delaying $t_\textrm{max}$ and resulting in a slight recovery of $E_\textrm{max}$. In contrast, the $t_\textrm{max}$ achieved with various $\Omega_{\mathrm{45}}$ demonstrates a relatively high stability.}

\begin{figure}
	\includegraphics[width=0.48\textwidth]{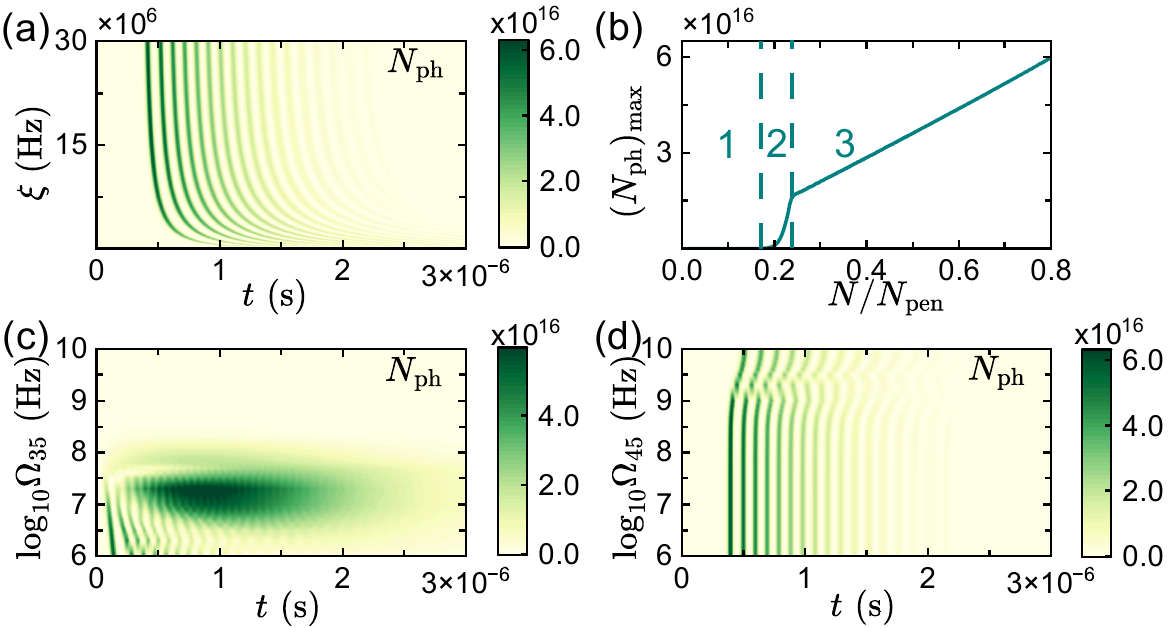}
	\caption{\label{fig4} The effects of the pump intensity $\xi$ (a) and driving fields  $\Omega_{\mathrm{35}}$ (c) and $\Omega_{\mathrm{45}}$ (d) on the dynamics of the work extraction revealed by the time-dependent microwave photon number in the cavity. (b) The dependence of the maximum photon number $(N_{\mathrm{ph}})_{\mathrm{max}}$ on the size of the quantum battery $N$ under  $\xi=6.2\times10^7$ Hz. Three different scaling regions are labelled.}
\end{figure}

By analyzing the time-dependent microwave photon number,  $N_{\mathrm{ph}}=\langle\hat{a}^{\dagger} \hat{a}\rangle$, the characteristics of the work extraction are shown in Fig.\hyperref[fig4]{4}. 
Without the driving fields, there is a latency for the energy to be extracted which ranges from about 500 ns to 1 $\mu$s depending on the pump intensity. The stronger pumping charges the battery faster to reach the maximum for the work extraction and shorten the latency time which coincides with the occurrence of the first energy dip (see Fig.\hyperref[fig4]{2(b)}) following $E_\textrm{max}$ as shown in Fig.\hyperref[fig4]{4(a)}.  
The oscillatory feature of $N_{\mathrm{ph}}$ is the Rabi oscillation that implies the microwave emission is from the collective Dicke states\cite{Breeze2017} formed by the pentacene triplet spin ensemble efficiently coupled to the microwave photons in the cavity. 

Fig.\hyperref[fig4]{4(b)} reveals the effect of battery size in work extraction. When $N$ is low, the small stored energy cannot compensate the energy dissipation of the load and the generated microwave photons are negligible (region $1$). Once $N$ reaches $\sim2\times10^6$ (region $2$), the stimulated emission starts to play a role for energy release and the photon number increases nonlinearly. When $N$ is high enough, the stimulated emission dominates the dynamics and the number of microwave photons increases linearly (region $3$). The results correspond to a phase transition observed in various masers\cite{doi:10.1073/pnas.1311866110, PhysRevA.50.4318}.

Additionally, the driving fields can dramatically modify the dynamics of the work extraction. \cor{Fig.\hyperref[fig4]{4(c)} shows that $\Omega_{\mathrm{35}}$ can significantly shorten the latency for the surge of the microwave photons before $10^7  $ Hz. However, in the subsequent phase, it increases both the delay time and the duration of the emission peak. When the value exceeds approximately $10^8 $ Hz, the driving field $\Omega_{\mathrm{35}}$ inhibits the generation of microwave photons. As depicted in Fig.\hyperref[fig4]{4(d)}, the influence of parameter $\Omega_{\mathrm{45}}$ on the emission peak is not significant before the value of $10^9 $ Hz. Beyond this threshold, an increase in $\Omega_{\mathrm{45}}$ leads to a delay in the arrival of the emission peak. These observational results are in close agreement with the dependence of $t_\textrm{max}$ and $E_\textrm{max}$ on $\Omega_{\mathrm{35}}$ and $\Omega_{\mathrm{45}}$ shown in Fig.\hyperref[fig4]{3}.}

In summary, we explore the role of metastability in quantum thermodynamics by investigating a solid-state quantum battery based on pentacene molecules. The metastability mitigates energy fluctuations during charging and offers a $\mu$s-level energy-storage lifetime and efficient work extraction. Notably, the pentacene quantum battery shows scaling behaviors exceeding the typical limits and the controllable charging performance. For practical applications, the pentacene quantum batteries can power a microwave quantum oscillator for coherent microwave emission. We note, due to the metastability, the quantum batteries can be recycled for phosphorescence generation, i.e., second-stage work extraction, until the pentacene molecules decay to their ground state. 
This work highlights the importance of metastability in advancing the field of quantum batteries and establishes a foundational framework for developing pure quantum devices powered by these batteries.



\begin{acknowledgments}
\textit{Acknowledgements.---}The work is supported by the National Natural Science Foundation of China (Grant No. 12204040), and the Beijing Institute of Technology Research Fund Program for Young Scholars (Grant No. XSQD-6120230016).
\end{acknowledgments}

\section{}
\subsection{}
\subsubsection{}

\end{document}